\documentclass[onecolumn,showpacs,amsmath,amssymb,aps,prb]{revtex4}

\usepackage{graphicx} 
\usepackage{bm} 
\usepackage{epsfig}

\begin{document}

\title{Transverse phase-locking in fully frustrated Josephson junction
arrays: a new type of fractional giant steps}
\author{Ver\'{o}nica I. Marconi}
\affiliation{The Abdus Salam  International Centre for Theoretical
Physics, I-34014, Trieste, Italy}
\author{Alejandro B. Kolton}
\author{Daniel Dom\'{\i}nguez}
\affiliation{Centro At\'{o}mico Bariloche,
 8400 San Carlos de Bariloche, R\'{\i}o Negro, Argentina.}
\author{Niels Gr{\o}nbech-Jensen}
\affiliation{Department of Applied Science,
 University of California, Davis, California 95616\\
and Computational Research Division, Lawrence Berkeley National Laboratory, Berkeley, California 94720.}

\today

\begin{abstract}
We study, analytically and numerically, phase locking of driven vortex
lattices in fully-frustrated Josephson junction arrays at zero
temperature. We consider the case when an ac current is applied
{\it perpendicular} to a dc current. We observe phase locking, steps
in the current-voltage characteristics, with a dependence
on external ac-drive amplitude and frequency
qualitatively different from the Shapiro steps, observed when
the ac and dc currents are applied in parallel.
Further, the critical current increases with increasing transverse ac-drive
amplitude, while it decreases for longitudinal ac-drive.
The critical current and the phase-locked current step width, increase
quadratically with (small) amplitudes of the ac-drive. For larger amplitudes of
the transverse ac-signal, we find windows where the critical current is
hysteretic, and windows where phase locking is suppressed due to dynamical
instabilities. We
characterize the dynamical states around the phase-locking
interference condition in the $IV$ curve with voltage noise, Lyapunov
exponents and Poincar\'e sections. We find that zero temperature phase-locking
behavior in large fully frustrated arrays is well described by an
effective four plaquette model.
\end{abstract}


\maketitle

\section{INTRODUCTION}
Phase locking phenomena are found in a wide variety
of nonlinear driven systems in condensed matter physics.\cite{bak}
It takes place when an internal frequency of
the system locks to a rational multiple of the frequency
of an external ac-drive. A simple example of this is
the case of an overdamped particle moving in a tilted
washboard potential, where the frequency of motion
of the particle over the periodic potential can be locked
to multiples of the frequency of a superimposed ac force for
a finite range of the dc force (tilt of the washboard).
Since the internal (washboard) frequency
is proportional to the mean velocity of
the particle, phase locking results in a constant mean
velocity for a certain range of dc-force curve when
the interference condition
is satisfied. A particularly well known realization of this effect is Shapiro
steps\cite{shapiro} in the dc current-voltage ($IV$) characteristics of a
single small area Josephson
junction driven by a time periodic current.
Within the washboard analogy outlined above, a simple analysis
provides expressions for the appearance of Shapiro steps at
specific voltages \cite{barone}
corresponding to integer multiples of the driving frequency.

Driven systems with many degrees of freedom can also exhibit phase locking.
This has attracted broad scientific and  technological interest since
phase-locking in complex systems can either be induced by collective
effects, providing for a low dimensional interpretation of the phenomenon,
or itself induce collective (low dimensional) behavior in the complex system.
Phase-locking experiments have provided
information about such dynamical response of non-equilibrium collective
states, where dimensionality, thermal fluctuations, quenched disorder, and the
magnitude of  external fields can be very relevant. A particular well known
example is the large Josephson junction array (JJA), with $N \times N$
junctions, driven by an external current $(I_{dc}+I_{ac} \cos \Omega t){\hat x}$
with frequency $\Omega$ and with an applied magnetic field density 
$f= H a^2/\Phi_0$, where  $H$  is  the magnetic field, $a$ the lattice period of
the Josephson array and $\Phi_0$  the quantum of flux.
{\it Giant} Shapiro steps  at voltages $V_n= N n\hbar \Omega/
2e$, $n$ being an integer, have been observed experimentally
in  zero magnetic field ($f=0$). \cite{leeman}
{\it Fractional giant} Shapiro steps at voltages $V_{n,q}=Nn\hbar\Omega/2eq$,
 were observed experimentally \cite{expfgs,expshap}
for strongly commensurate magnetic fields, $f=p/q$, with $p, q$
integers, and extensively investigated in numerical
simulations.\cite{lee,simff,fourp,octav,ciria}
Also, {\it subharmonic} giant Shapiro steps at voltages
$V_{n,m}=Nn\hbar\Omega/2em$
were observed experimentally \cite{expshap}
for zero magnetic field ($f=0$),  and 
attributed to the nucleation
of complex  collective dynamical states \cite{acvs,giov} induced
by disorder or inductance effects. Shapiro-like phase-locking is also 
observed in the case of  driven vortex lattices in bulk superconductors with
two-dimensional periodic  pinning arrays,
as recently reported both experimentally \cite{vanlook} and
theoretically.\cite{reichhardt} 
Also superconductors, where vortices are
driven over a one-dimensional 
potential generated by thickness modulations,\cite{martinoli} or
are confined to move through mesoscopic channels, \cite{kokubo}
show Shapiro-like phase-locking. Moreover,
systems with many degrees of freedom in the presence of  quenched disorder can
also exhibit phase-locking when  there is a dynamically induced periodicity,
like charge  density waves \cite{gruner} and vortex lattices in
superconductors  with random pinning.\cite{fiory,harris,kolton}

In the phase-locking examples mentioned above, the ac-drive is applied
parallel to the dc-drive. However, it was recently shown that a
different type of phase-locking, distinct from
Shapiro phase-locking, is
possible in vortex lattices if the ac-force is applied
{\it perpendicular} to the dc-force.\cite{reichhardt2,kolton3}
In this case the interference effect is due to an effective {\it parametric}
ac drive in the longitudinal direction, which is induced by the
transverse ac drive. In several systems, like charge density waves
or single degree of freedom systems ({\it e.g.}, the single Josephson
junction), the dynamical variables are such that
perturbations or displacements can be induced in only one direction
({\it i.e.}, the displacement field is a scalar).
An important characteristic of vortex lattices in superconductors
is that the displacement field is two dimensional.  In
particular, the behavior of displacements in the direction perpendicular to
the driving force shows phenomena like a transverse critical current
\cite{koshelev,moon,olson,marconi} and a transverse freezing transition
\cite{koshelev,kolton2} at high velocities. Phase locking in ac-driven vortex
lattices, where the ac and dc forces are perpendicular, arises as a direct
consequence of the nonlinear coupling between the two directions of motions.

Transverse phase-locking has been reported for vortex lattices moving in
rectangular or triangular pinning arrays \cite{reichhardt2} as well as in arrays
of randomly distributed  pinning centers.\cite{kolton3} In this  paper we
investigate the possibility of transverse phase locking in a two dimensional (2D)
fully frustrated JJA, where the average of the external magnetic field corresponds
to one half flux quantum per plaquette, $f=1/2$. 
This system has several attractive properties. The presence of a
magnetic field ($f \neq 0$) breaks the axial symmetry in the direction of the
bias  current, and 2D-cooperative behavior may come into play. This leads
to the well-known fractional giant Shapiro steps
\cite{expfgs,expshap,lee,simff,fourp,octav,ciria} induced
by a longitudinal ac current. It also, as we will demonstrate in this paper,
allows for {\it transverse} phase locking when
the ac current is perpendicular to the dc current, since the two directions
of motion become coupled. Non-equilibrium dynamical phases for fully frustrated JJAs
driven by a dc current have previously been studied.\cite{niels,marconi2}
Phase locking can be used to characterize temporal order in the different
dynamical phases of the JJA at high velocities by their ac-response, as was
done in Ref.~\onlinecite{harris,kolton} for bulk superconductors.

Here, we report transverse phase locking steps in the
$IV$ characteristics, similar to
the longitudinal giant Shapiro steps, but with very
different characteristic dependencies on external ac-drive amplitude
$I_{ac}$ and frequency $\Omega$. The critical depinning
current in the system with transverse ac drive
is {\it larger} than the critical current of the
dc driven system. For $I_{ac}/\Omega \ll 1$ the
depinning critical current $I_c$ and the phase-locked
step width $\Delta S_1$ for $V=\Omega \hbar/2e$ increase
quadratically with $I_{ac}$. For $I_{ac}/\Omega > 1$ we find
windows of $I_{ac}/\Omega$
where depinning is hysteretic and the periodic phase-locked dynamical
states become unstable. We characterize
the dynamical states around the phase-locking interference condition
in the $IV$ curve with the voltage noise, Lyapunov exponents and
Poincar\'e sections. We find that zero temperature phase-locking
behavior in large fully frustrated arrays is well described by an
effective four plaquette model.

The remainder of this paper is outlined as follows. In section II
we present the model used for simulating the dynamics of the
fully frustrated JJA. In section III we develop an analytical
framework for predicting critical current and phase-locking
properties of the fully frustrated JJA. Section IV presents
simulated transverse phase-locking
steps in typical $IV$ curves obtained from
an effective four plaquette model for the JJA. We calculate numerically
the dependence of the critical current and the magnitude of phase-locking steps
with the amplitude $I_{ac}$ and frequency $\Omega$ of the external ac-drive.
The results obtained are analyzed in more detail by studying
voltage noise, Lyapunov exponents and Poincar\'e sections for the
dynamical states around the phase-locking interference
condition. We also compare numerical simulations results
for large arrays with those obtained using the effective four plaquette model.
The discussions and conclusions of the investigation are presented in
sections IV and V.

\section{MODEL}

\begin{figure}[tbp]
\centerline{\epsfxsize=8.5cm \epsfbox{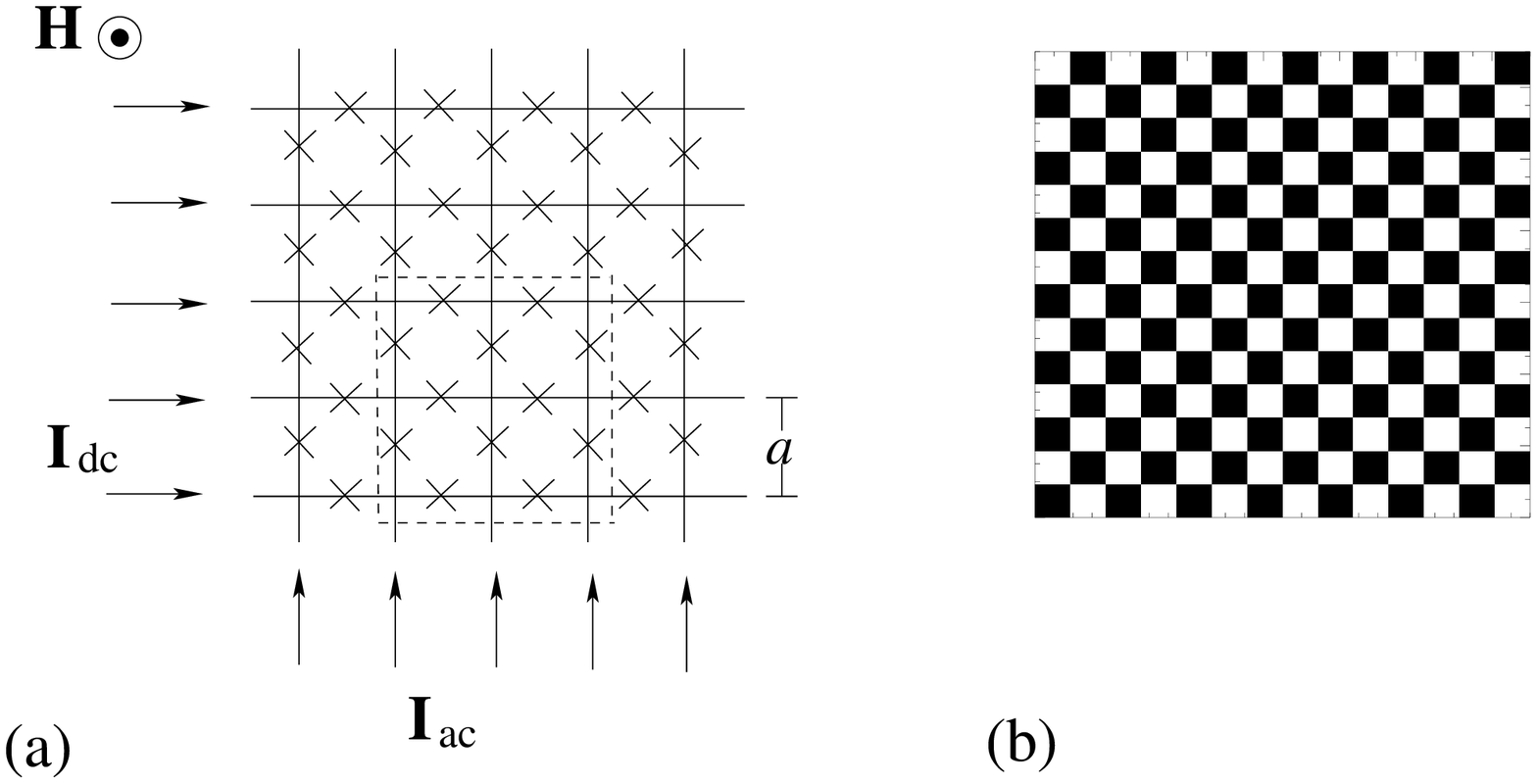}}
\caption{
(a) Schematic JJA showing driving currents, and the repeated two-junction
by two-junction superlattice unit cell in the ground state. (b) Ground
state ``checkerboard'' vorticity for $f= H a^2 / \Phi_0 = 1/2$. Black
squares represent plaquettes with one vortex, white squares represent
plaquettes without vortices.}
\label{fig:fig1}
\end{figure}

We study a current driven JJA with an ac current perpendicular to the dc current,
as shown in figure \ref{fig:fig1}a. A magnetic field, $H$,  is applied such that
half a flux quantum, $f=Ha^2/\Phi_0=1/2$ with $a^2$ being the area of a plaquette
and $\Phi_0=h/2e$ being the flux quantum, penetrates each plaquette; corresponding to
the fully frustrated XY model,\cite{xyff, mon}
where the ground state is a ``checkerboard'' vortex lattice, in which a
vortex (flux quantum) penetrates every other square grid (see Fig.~\ref{fig:fig1}b).
In such ground state, current and phase differences in the junctions are described by a
repeated two-junction by two-junction ($2\times2$ plaquette) superlattice unit cell.

\begin{figure}[tbp]
\centerline{\epsfxsize=5.5cm \epsfbox{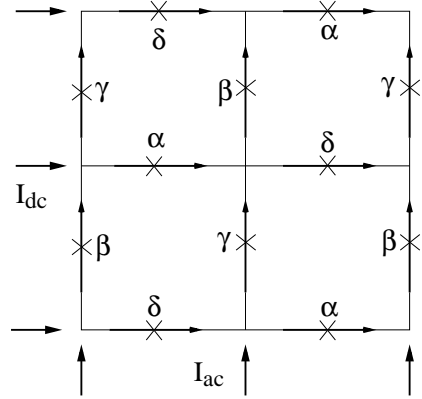}}
\caption{Square four-plaquette model, being $\alpha$, $\beta$,
$\gamma$, $\delta$, the gauge-invariant phase differences.}
\label{fig:fig2}
\end{figure}

Numerical simulations \cite{simff,lee} of large driven arrays suggest that
this spatial periodicity is preserved when the dynamics is phase-locked to an
external ac-perturbation applied in parallel to the dc force.
We will show later, in section IV, that this is also a good
approximation when the ac current is {\it perpendicular} to
the dc current. We will therefore consider the simple system of a $2\times2$
superlattice unit cell of the array and the associated
gauge-invariant phase differences
in a field of $f=1/2$, with the dc current (per plaquette)
$I_{dc}$ perpendicular to an ac current (per plaquette) $I_{ac} \cos(\Omega t)$,
as shown in Fig.~\ref{fig:fig2}.
Flux quantization, total current conservation at the central node, and the
applied total currents in the two directions give the following governing equations,
\begin{eqnarray}
\alpha+\beta-\delta-\gamma & = & \pi(1+2n) \label{eq:Eq_1} \\
\dot{\beta}+\dot{\gamma}+\sin\beta+\sin\gamma & = & 2I_{ac}\cos\Omega{t} \label{eq:Eq_2} \\
\dot{\alpha}+\dot{\delta}+\sin\alpha+\sin\delta & = & 2I_{dc} \label{eq:Eq_3} \\
\dot{\alpha}+\dot{\gamma}-\dot{\beta}-\dot{\delta}+ \sin\alpha &+&
\sin\gamma - \sin\beta-\sin\delta \; = \; 0 \; , \label{eq:Eq_4}
\end{eqnarray}
where $n$ is an integer, $t$ is the normalized time
in units of $t_0=2eI_0 R_N/\hbar$, $R_N$ being the normal state single junction resistance,
$\Omega$ is the normalized frequency in units
of $1/t_0$, $I_{ac}$ and $I_{dc}$ are the normalized external currents in
units of the single junction critical current $I_0$.
This model was introduced by Benz {\it et al.} \cite{ccff} (for $I_{ac}=0$)
to study the dc current-voltage curve of the $f=1/2$ array. They obtained
analytically that the critical current per junction
of this model is $I_c=(\sqrt{2}-1)I_0$. \cite{ccff}
The same model was later used in Ref.~\onlinecite{fourp} to study the (longitudinal)
Shapiro steps. Since the analysis done in the work of Refs.~\onlinecite{ccff,fourp}
did not include the transverse ac current with accompanying transverse voltage
drop, an additional constraint of $\beta=\delta$ was implied, reducing the model system
to two dynamical degrees of freedom. In contrast, our model system of a four plaquette
unit cell consists of three effective dynamical variables. We
calculate the instantaneous longitudinal $V_x$ and transverse $V_y$ (normalized)
voltages per junction as,
\begin{eqnarray}
V_x & = & (\dot \alpha+\dot \delta)/2 \label{eq:Eq_5}  \\
V_y & = & (\dot \beta+\dot \gamma)/2 \; , \label{eq:Eq_6}
\end{eqnarray}
and the $IV$ characteristics,
$v_x=\langle V_x \rangle$ as a function of $I_{dc}$, where $\langle\cdots\rangle$ is
a time average.
The total longitudinal voltage $v_T$ for an $N \times N$ array, built
with this $2 \times 2$ superlattice unit cell, is $v_T = N v_x$.
When vortices move with a mean velocity $u$ in such $2a\times 2a$
superlattice structure, we can obtain
the normalized voltage $v_x$ using the relation
$2 \pi u/ 2a = v_x$, where $a$ is the array periodicity (see Fig.~\ref{fig:fig1}a).
The intrinsic washboard frequency for vortices moving with velocity $u$
in the periodic potential of the JJA is $\omega_0= 2\pi u /a$.
Phase-locking in the longitudinal direction is obtained
when the frequency $\Omega$ of the ac drive locks to a rational multiple of the
intrinsic washboard frequency. For  the $n$-th harmonic this 
corresponds to $\omega_0=n\Omega$, {\it i.e.} $2\pi u/a = n \Omega$.
This leads to phase-locking at voltages $V_{n,2}=(n/2)\Omega$ for
fully frustrated JJA.
In general, for $f=p/q$, the ground state has $qa\times qa$
superlattice structure, therefore the voltage for vortices moving with
velocity $u$ is $2 \pi u/ qa = v_x$,
 and phase-locking for the $n$-th harmonic 
 is obtained at voltages $V_{n,q}=(n/q)\Omega$.
 This is the condition for the so-called ``fractional giant Shapiro steps''
\cite{expfgs,expshap,lee,simff,fourp,octav,ciria}.

\section{Phase-Locking and Critical Current Analysis}

We will in this section assume that the dynamics of the system is represented
by the simple two-plaquette degrees of freedom as shown in figure \ref{fig:fig2}.
We will apply the following linear transformation of the phase-variables
of Eqs.~(\ref{eq:Eq_1})-(\ref{eq:Eq_4}):
\begin{eqnarray}
\Phi_x & = & \frac{\alpha+\delta}{2} \label{eq:Eq_7}\\
\Psi_x & = & \frac{\alpha-\delta}{2} \label{eq:Eq_8}\\
\Phi_y & = & \frac{\beta +\gamma}{2} \label{eq:Eq_9}\\
\Psi_y & = & \frac{\beta -\gamma}{2} \; . \label{eq:Eq_10}
\end{eqnarray}
With these variables we can represent the constraint ($f=\frac{1}{2}$)
of Eq.~(\ref{eq:Eq_1}) as $\Psi_x+\Psi_y=\frac{\pi}{2}$, and thereby write the relevant
three degrees of freedom in either of the two following forms, eliminating
$\Psi_x$:
\begin{eqnarray}
\dot{\Phi}_x+\sin{\Psi_y}\sin{\Phi_x} & = & I_{dc} \label{eq:Eq_11} \\
\dot{\Phi}_y+\cos{\Psi_y}\sin{\Phi_y} & = & I_{ac}\cos\Omega{t} \label{eq:Eq_12} \\
2\dot{\Psi}_y-\cos{\Phi_x}\cos{\Psi_y}+\cos{\Phi_y}\sin{\Psi_y} & = & 0 \; , \label{eq:Eq_13}
\end{eqnarray}
or eliminating $\Psi_y$:
\begin{eqnarray}
\dot{\Phi}_x+\cos{\Psi_x}\sin{\Phi_x} & = & I_{dc} \label{eq:Eq_14} \\
\dot{\Phi}_y+\sin{\Psi_x}\sin{\Phi_y} & = & I_{ac}\cos\Omega{t} \label{eq:Eq_15} \\
2\dot{\Psi}_x+\cos{\Phi_x}\sin{\Psi_x}-\cos{\Phi_y}\cos{\Psi_x} & = & 0 \; . \label{eq:Eq_16}
\end{eqnarray}
The normalized voltages are given by, $V_x=\dot{\Phi}_x$ and $V_y=\dot{\Phi}_y$.
We will in the spirit of the usual Shapiro analysis assume that
\begin{eqnarray}
\Phi_y & = & \frac{I_{ac}}{\Omega}\sin{\Omega t} \; , \label{eq:Eq_17}
\end{eqnarray}
which is the solution to Eqs.~(\ref{eq:Eq_12}) and (\ref{eq:Eq_15}) for large $I_{ac}$ and $\Omega$.

\subsection{Critical Current}

We will here look at Eqs.~(\ref{eq:Eq_14}) and (\ref{eq:Eq_16}). Assuming
$\Phi_x=\Phi_x^{(0)}$ and $\Psi_y=\Psi_y^{(0)}+\varepsilon(t)$, where
$\Phi_x^{(0)}$ and $\Psi_y^{(0)}$ are constants and
$\big|\varepsilon(t)\big|\ll1$, the static contribution to equation (\ref{eq:Eq_16}) is:
\begin{eqnarray}
\cos{\Phi_x^{(0)}}\sin{\Psi_y^{(0)}} & = & J_0\left(\frac{I_{ac}}{\Omega}\right)\cos{\Psi_y^{(0)}} \label{eq:Eq_18} \\
\Rightarrow \; \; \tan{\Psi_y^{(0)}} & = & \frac{J_0\left(\frac{I_{ac}}{\Omega}\right)}{\cos{\Phi_x^{(0)}}} \; , \label{eq:Eq_19}
\end{eqnarray}
where $J_n$ is the $n$'th order Bessel function of the first kind.
Inserting this into the static part of equation (\ref{eq:Eq_14}) yields
\begin{eqnarray}
I_{dc} & = & \cos\left\{\tan^{-1}\left[\frac{J_0\left(\frac{I_{ac}}{\Omega}\right)}{\cos{\Phi_x^{(0)}}}\right]\right\}\sin{\Phi_x^{(0)}} \; . \label{eq:Eq_20}
\end{eqnarray}
This expression provides a unique relationship between the constant phase,
$\Phi_x^{(0)}$, and the dc current, $I_{dc}$.
However, for increasing $I_{dc}$, there exists a critical value,
$I^{\uparrow}_{c}$, for which no real $\Phi_x^{(0)}$ can satisfy Eq.~(\ref{eq:Eq_20}).
This value is given by
\begin{eqnarray}
I^{\uparrow}_{c} & = & {\rm max}\left[\cos\left\{\tan^{-1}
\left[\frac{J_0\left(\frac{I_{ac}}{\Omega}\right)}
{\cos{\Phi_x^{(0)}}}\right]\right\}\sin{\Phi_x^{(0)}}\right] \; ,
 \label{eq:Eq_21}
\end{eqnarray}
which is the predicted critical dc current for static states
($v_x=\langle\dot{\Phi}_x\rangle=0$). We notice that the identical expression for the
critical current, Eq.~(\ref{eq:Eq_21}), can be obtained from the {\it anisotropic}
dc driven system,
\begin{eqnarray}
\alpha+\beta-\delta-\gamma & = & \pi(1+2n) \label{eq:Eq_22} \\
\dot{\beta}+\dot{\gamma}+\Gamma   \sin\beta+\Gamma   \sin\gamma & = & 0 \label{eq:Eq_23} \\
\dot{\alpha}+\dot{\delta}+\sin\alpha+\sin\delta & = & 2I_{dc} \label{eq:Eq_24} \\
\dot{\alpha}+\dot{\gamma}-\dot{\beta}-\dot{\delta}+ \sin\alpha &+&
\Gamma   \sin\gamma - \Gamma   \sin\beta-\sin\delta \; = \; 0 \; , \label{eq:Eq_25}
\end{eqnarray}
where the anisotropy, $\Gamma$ (suppression of transverse critical current),
is given by the standard Shapiro \cite{barone} critical current,
$\Gamma=J_0(\frac{I_{ac}}{\Omega})$.

\subsection{Phase-Locking}
We will here use equations (\ref{eq:Eq_11})-(\ref{eq:Eq_13}), since
$\langle {\Psi}_y \rangle = 0 ({\rm mod}\pi)$ for
$\langle\dot{\Phi}_x\rangle \neq 0$ provides for a simple description of
the dynamics.
We will assume the ansatz, $\Phi_x=\Phi_x^{(0)}+\Omega t$
and $\Psi_y=A\sin{\Omega t}+B\cos{\Omega t}$. The equation for $\Psi_y$, (\ref{eq:Eq_13}), now reads:
\begin{eqnarray}
2\dot{\Psi}_y-\cos(\Phi_x^{(0)}+\Omega t)\cos{\Psi_y}+\sin{\Psi_y}\sum_{k}J_k\left(\frac{I_{ac}}{\Omega}\right)\cos{k\Omega t} & = & 0 \; , \label{eq:Eq_26}
\end{eqnarray}
where we will use the approximations:
$\cos{\Psi_y}\approx J_0\left(\sqrt{A^2+B^2}\right)$ and $\sin{\Psi_y}\approx{\Psi_y}$.
With the ansatz for $\Psi_y$ above, this equation has no static component. The
time varying component, at frequency $\Omega$, yields the coefficients $A$ and $B$
\begin{eqnarray}
A_1 & = & A_0J_0\left(\sqrt{A_0^2+B_0^2}\right) \; = \; \frac{2\Omega\cos{\Phi_x^{(0)}}-\left[J_0\left(\frac{I_{ac}}{\Omega}\right)+J_2\left(\frac{I_{ac}}{\Omega}\right)\right]\sin{\Phi_x^{(0)}}}{4\Omega^2+J_0^2\left(\frac{I_{ac}}{\Omega}\right)-J_2^2\left(\frac{I_{ac}}{\Omega}\right)} J_0\left(\sqrt{A_0^2+B_0^2}\right) \label{eq:Eq_27} \\
B_1 & = & B_0J_0\left(\sqrt{A_0^2+B_0^2}\right) \; = \; \frac{2\Omega\sin{\Phi_x^{(0)}}+\left[J_0\left(\frac{I_{ac}}{\Omega}\right)-J_2\left(\frac{I_{ac}}{\Omega}\right)\right]\cos{\Phi_x^{(0)}}}{4\Omega^2+J_0^2\left(\frac{I_{ac}}{\Omega}\right)-J_2^2\left(\frac{I_{ac}}{\Omega}\right)} J_0\left(\sqrt{A_0^2+B_0^2}\right) \; , \label{eq:Eq_28}
\end{eqnarray}
where $A_0$ and $B_0$ are the solutions for
$J_0\left(\sqrt{A_0^2+B_0^2}\right)=1$. Thus, the solution $(A,B)=(A_0,B_0)$
is correct up to order $\Omega^{-1}$ and $(A,B)=(A_1,B_1)$ is correct
up to $\Omega^{-2}$.
Inserting this solution $(A_1,B_1)$
for $\Psi_y$ into the $\Psi_y$-linearized equation (\ref{eq:Eq_11})
gives the static properties:
\begin{eqnarray}
J_0\left(\sqrt{A_0^2+B_0^2}\right) \frac{1}{2}\left(A_0\cos{\Phi_x^{(0)}}+B_0\sin{\Phi_x^{(0)}}\right) & = & I_{dc}-\Omega \; \; \Rightarrow \label{eq:Eq_29} \\
J_0\left(\sqrt{A_0^2+B_0^2}\right) \frac{1}{2}
\frac{2\Omega-J_2\left(\frac{I_{ac}}{\Omega}\right)\sin{2\Phi_x^{(0)}}}
{4\Omega^2+J_0^2\left(\frac{I_{ac}}{\Omega}\right)-J_2^2\left(\frac{I_{ac}}{\Omega}\right)} & \approx & 
\Delta I_1 - \frac{1}{2}\Delta S_1 \sin{2\Phi_x^{(0)}} \nonumber \\
 & =  & I_{dc}-\Omega \; . \label{eq:Eq_30} 
\end{eqnarray}
Thus, the locking range for the this step can be found to second order in $|\Psi_y|$.
The dominant part of this expression for the range in phase-locking yields:
\begin{eqnarray}
\Delta S_1 & = & \frac{\Big|J_2\left(\frac{I_{ac}}{\Omega}\right)\Big|}
{4\Omega^2+J_0^2\left(\frac{I_{ac}}{\Omega}\right)-J_2^2\left(\frac{I_{ac}}{\Omega}\right)}
\left(1-\frac{3\Omega^2+\frac{1}{4}\left(J_0^2\left(\frac{I_{ac}}{\Omega}\right)+J_2^2\left(\frac{I_{ac}}{\Omega}\right)\right)}
{\left(4\Omega^2+J_0^2\left(\frac{I_{ac}}{\Omega}\right)-J_2^2\left(\frac{I_{ac}}{\Omega}\right)\right)^2}
\right) \; . \label{eq:Eq_31}
\end{eqnarray}
The expression displays quadratic growth of the phase-locked step size
for small $I_{ac}$. This is consistent with the particle (pancake) model
results \cite{reichhardt2} for vortices in rectangular pinning arrays, and
it is different from the known longitudinal (Shapiro) phase-locking of
JJAs.\cite{leeman,expfgs,expshap,lee}

\begin{figure}[tbp]
\centerline{\epsfxsize=8.5cm \epsfbox{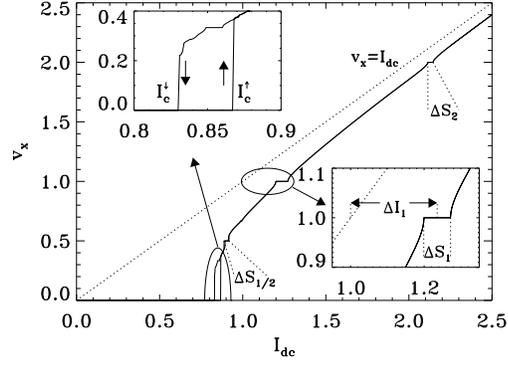}}
\caption{$IV$ curve of a $2\times2$ JJA for $I_{ac}=2.3$ and $\Omega=1$,
 showing transverse phase locking and critical currents. Inset shows a
 detail of the hysteresis around the critical current.}
\label{fig:fig3}
\end{figure}

In addition to the range of phase-locking, $\Delta S_1$, equation (\ref{eq:Eq_30}) provides
information about the offset, $\Delta I_1$, of the phase-locked step relative to
the Ohmic (linear) curve (see inset in figure \ref{fig:fig3}).
The offset is given by the part of the equation that does {\it not}
depend on $\Phi_x^{(0)}$. From Eq.~(\ref{eq:Eq_30}) we have:
\begin{eqnarray}
\Delta I_1 & = & \frac{\Omega}{4\Omega^2+J_0^2\left(\frac{I_{ac}}{\Omega}\right)-J_2^2\left(\frac{I_{ac}}{\Omega}\right)}\left(1-\frac{1}{4}\frac{4\Omega^2+J_0^2\left(\frac{I_{ac}}{\Omega}\right)+2J_2^2\left(\frac{I_{ac}}{\Omega}\right)}{\left(4\Omega^2+J_0^2\left(\frac{I_{ac}}{\Omega}\right)-J_2^2\left(\frac{I_{ac}}{\Omega}\right)\right)^2}\right) \; . \label{eq:Eq_32}
\end{eqnarray}

Expressions (31) and (32) provide a second order (in $\Omega^{-1}$) description of
the phase-locking step magnitude and location as a function of the system parameters,
$\Omega$ and $I_{ac}$, for large $\Omega$.

\section{RESULTS AND DISCUSSION}

We will here show the results of numerical simulations of the system analyzed
in the previous section. The simulations are conducted with numerical parameters
corresponding to the model parameters, using a fourth order Runge-Kutta method
such that the normalized time step
typically is no larger than 1\% of the period of the driving frequency, and
often smaller. Since we are mostly concerned with dc $IV$ characteristics,
we choose to acquire data for averaging over many ac-periods of motion (typically
10$^2$-10$^3$) after a sufficient initial time of interval allowed for transient
behavior. Simulated $IV$ characteristics are obtained by performing the
necessary
averages as described, and then changing the dc current, $I_{dc}$, slightly to
acquire the next point on the $IV$ curve. All simulations are conducted for the fully
frustrated case of $f=\frac{1}{2}$.

Figure \ref{fig:fig3} shows a simulated $IV$ characteristic, $v_x$ as a function of
$I_{dc}$, simulated for $\Omega=1$ and
$I_{ac}=2.3$, for the simple four plaquette model showed in figure \ref{fig:fig2}, described
by the equations (\ref{eq:Eq_1})-(\ref{eq:Eq_4}).
As is obvious from the figure, we obtain clear signatures of critical
current(s) and phase-locked steps. Specifically, we observe the $\Delta S_1$ step
at $v_x=\Omega$, and steps $\Delta S_{\frac{1}{2}}$ and $\Delta S_2$ at
$v_x=\frac{1}{2}\Omega$ and $v_x=2\Omega$, respectively. We observe
a critical current larger than the previously predicted
\cite{ccff} value of $I_c=\sqrt{2}-1 \approx 0.41$ for a fully frustrated
dc-driven system. The characteristics of this plot are very similar to
the behavior observed in JJA with parallel ac $+$ dc drives,
obtained both by simulations \cite{simff, octav} and  experiments,\cite{expfgs}
as well as  analytically for the 
four plaquette model.\cite{fourp}

Comparing simulations of the full-RSJ dynamical 
equations \cite{marconi,marconi2} for
different large arrays ($N \times N$ junctions) we will later (below)
demonstrate that the
simple model of a $2\times2$ array gives very good description of the JJA
dynamics. However, we will first compare the predictions of the analytical
treatment of the previous section to numerical results.

\subsection{Critical Current and Phase-Locking for $\Omega\ge1$}
In order to verify the simple theory for critical current and
phase-locking behavior developed above, we have conducted numerical
simulations of the four plaquette system described by Eqs.~(\ref{eq:Eq_1})-(\ref{eq:Eq_4})
for $1\le\Omega$.

\begin{figure}[tbp]
\centerline{\epsfxsize=8.5cm \epsfbox{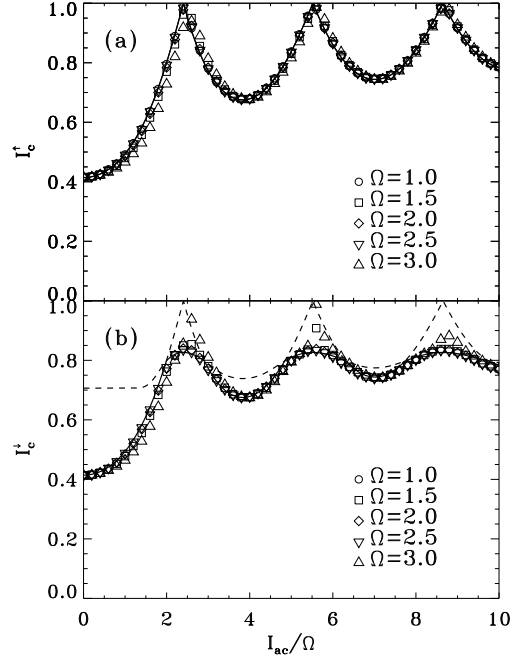}}
\caption{Critical current, $I_c$, of a $2\times2$ JJA vs ac-amplitude and
frequency, $I_{ac}/\Omega$ for high frequencies, $\Omega \ge 1$.
The hysteresis of the critical current is demonstrated by (a) the critical
current, $I_{c}^\uparrow$, switching from zero to non-zero voltage state, and
(b) the critical current, $I_{c}^\downarrow$, switching from non-zero to zero
voltage states. See Figure 3. Markers are results of numerical simulations and
lines are the corresponding predictions: (a) equation (\ref{eq:Eq_21});
(b) dashed curve
is the maximum of equations (\ref{eq:Eq_21}) and (\ref{eq:Eq_37}),
while the solid curve is the
minimum of the two. The critical current, $I_{c}^\downarrow$, is predicted to
follow the solid curve.
}
\label{fig:fig4}
\end{figure}

The first set of simulations are conducted to
investigate the critical current, $I_c$, as it is described by
Eq.~(\ref{eq:Eq_21}). This expression provides an estimate of the critical current,
$I_c^\uparrow$, for which the JJA switches from a zero-voltage state
($\langle V_x\rangle=0$ and $I_{dc}<I_c^\uparrow$) to a non-zero voltage
state. The simulations are conducted accordingly, starting the system
at rest for small $I_{dc}$ and slowly increasing the dc-bias until
non-zero average voltage is detected.
The results are shown in figure \ref{fig:fig4}a, where the solid curve represents
the expression, Eq.~(\ref{eq:Eq_21}), and the markers represent the simulation results
for several frequencies $1\le\Omega\le3$ as a function of the characteristic
ratio, $\frac{I_{ac}}{\Omega}$. The size of the markers are larger than the
error on the estimated critical current. It is obvious that the
agreement is very good for all simulated data sets, and we conclude that
the critical current, as given by Eq.~(\ref{eq:Eq_21}), is a relevant estimate for
$\Omega$ not smaller than 1.

Figure \ref{fig:fig4}b shows the critical current, $I_c^\downarrow$, evaluated
from numerical
simulations when the dynamical system switches from the non-zero voltage
state to the zero-voltage state (see inset in figure \ref{fig:fig3}).
We have here shown the results of numerical
simulations with markers as for figure \ref{fig:fig4}a, together with the solid curve
of Eq.~(\ref{eq:Eq_21}). However, it is clear from the figure that the critical
current, $I_c^\downarrow$, for decreasing dc-bias may be smaller than
for increasing dc-bias ($I_c^\uparrow\ge I_c^\downarrow$).
Since this is a multi-dimensional system, the critical current may be
hysteretic, such that decreasing the dc-current, $I_{dc}$, for non-zero
voltage states ($v_x=\langle\dot{\Phi}_x\rangle\neq0$) is subject to different
critical characteristics. One simple way of investigating this is to
assume a non-phase-locked state of voltage $v_x\neq0$,
such that $\Phi_x = v_xt$.
A primitive analysis can provide a hint to this hysteresis.

The critical current analysis of the previous section, is obviously a
critical current for a system operated at the $\langle\dot{x}\rangle=0$
branch of the $IV$-curve. We may instead
analyze what may happen for a {\it non-phase-locked}
$\langle\dot{\Phi}_x\rangle=v_x\neq0$ state. We will still assume (17) to be
an appropriate description of the transverse current.
However, equation (\ref{eq:Eq_13}) becomes (for small $|\Psi_y|$ and
with no resonance to $\Omega$):
\begin{eqnarray}
2\dot{\Psi}_y-\cos{v_xt} + J_0\left(\frac{I_{ac}}{\Omega}\right) \Psi_y & = & 0 \label{eq:Eq_33}\\
\Rightarrow \; \; \; \Psi_y & = & \frac{J_0\left(\frac{I_{ac}}{\Omega}\right)}{J_0^2\left(\frac{I_{ac}}{\Omega}\right)+4v_x^2}\cos{v_xt}+\frac{2v_x}{J_0^2\left(\frac{I_{ac}}{\Omega}\right)+4v_x^2}\sin{v_xt} \; . \label{eq:Eq_34}
\end{eqnarray}
Inserting the $\Psi_y$ solution into equation (\ref{eq:Eq_11}) yields the static component:
\begin{eqnarray}
v_x+\Big|J_1\left(\frac{2v_x}{J_0^2\left(\frac{I_{ac}}{\Omega}\right)+4v_x^2}\right)\Big| & = & I_{dc} \; , \label{eq:Eq_35}
\end{eqnarray}
where the second term on the left hand side is the result of the resonant mixing
between the propagation, $\langle\dot{\Phi}_x\rangle=v_x$ and the transverse
oscillation, $\Psi_y$. However, the overdamped dc driven pendulum equation
is also subject to the following simple relationship, \cite{barone}
\begin{eqnarray}
\sqrt{v_x^2+(I_c^{\downarrow})^2} & = & I_{dc} \; . \label{eq:Eq_36}
\end{eqnarray}
Combining the two expressions provides the relationship
\begin{eqnarray}
I_c^{\downarrow} & = & 
\sqrt{\left[v_x+\Big|J_1\left(\frac{2v_x}
{J_0^2\left(\frac{I_{ac}}{\Omega}\right)+4v_x^2}\right)
\Big|\right]^2-v_x^2} \; \; \le \; \; 0.826591 \; . 
\label{eq:Eq_37}
\end{eqnarray}
As we have indicated, the critical current, $I_c^\downarrow$ has a maximum value
at around 0.82 (for $J_0\left(\frac{I_{ac}}{\Omega}\right)=0$ and $v_x\approx0.33$).
Thus, we can argue that propagating ($\langle\dot{\Phi}_x\rangle\neq0$)
solutions may exist for $I_{dc}>I_c^{\downarrow}\approx0.82$, which provides
for a hysteretic $IV$ characteristic switching between zero and non-zero voltages
in the range $I_c^\downarrow\le I_{dc}\le I_c^\uparrow$, when
$I_c^\downarrow\le I_c^\uparrow$. Notice that when
$I_c^\downarrow >  I_c^\uparrow$, the relevant critical current for
both zero and non-zero voltage states must be $I_c^\uparrow$, since no static
states exist for $I_{dc}>I_c^\uparrow$. However, when $I_c^\downarrow<I_c^\uparrow$,
the actual critical current for switching into a zero-voltage state may be
anywhere in the interval $[I_c^\downarrow;I_c^\uparrow]$.

Figure \ref{fig:fig4} clearly indicates the hysteretic switching in the $IV$ characteristics
when $I_c^\downarrow\le I_{dc}\le I_c^\uparrow$, which is the case for small
$|J_0(\frac{I_{ac}}{\Omega})|$.

We note that the above rather primitive analysis of the hysteresis provides
a fairly good agreement with the results of numerical simulations. The results
of the analysis are not completely consistent with its assumptions in that
the resulting amplitude of $\Psi_y$ for the optimized $v_x\approx0.33$ is about
1.5, which is not a small number. However, a more detailed (nonlinear)
analysis of a single frequency representation of $\Psi_y$ yields quantitatively
similar and qualitatively identical results ($I_c^{\downarrow}\le0.77$) as the
above, and we therefore conclude that the simple explanation for hysteresis
presented here is relevant.

For $J_0\left(\frac{I_{ac}}{\Omega}\right)=0$ we can provide an explicit
approximate expression for $I_c^{\downarrow}$ by assuming the critical
current is given by the value of $v_x$ which optimizes $J_1\left(\frac{1}{2v_x}\right)$.
This leads to $v_x\approx1/3.6$, which inserted into the above equation yields
the optimized coordinates: $(I_c^{\downarrow},v_x)=(0.81,0.28)$.

It is noteworthy that we observe, as predicted by the expression for
$I_c^\uparrow$, $I_c(\Omega,I_{ac})$ to be {\it larger} than the dc-driven
system, $I_c(\Omega,I_{ac})\ge I_c(\Omega,0)\approx0.41$.
Hence, a transverse ac driving leads to an 
{\it enhancement} of the critical current.
This is contrary to the case with the ac-current parallel to the
dc-current, where the critical current is reduced, {\it i.e.}
 $I_c \le I_c(\Omega,0)\approx 0.41$.\cite{simff, octav, expfgs, fourp}

\begin{figure}[tbp]
\centerline{\epsfxsize=8.5cm \epsfbox{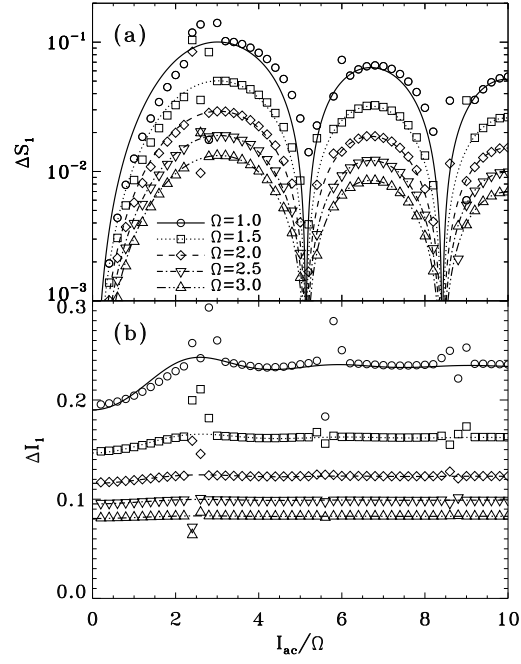}}
\caption{Phase-locking of a $2\times2$ JJA at $V_x=\Omega$.
Markers are results of numerical simulations
and lines are the corresponding predictions of equations (\ref{eq:Eq_31}) and
(\ref{eq:Eq_32}).
(a) Phase-locking range in dc current. (b) Offset of the phase-locked step relative
to the Ohmic curve. See Figure \ref{fig:fig3}.
}
\label{fig:fig5}
\end{figure}

The predicted range, $\Delta S_1$, in $I_{dc}$ of phase-locking, as given by
Eq.~(\ref{eq:Eq_31}), is investigated through simulations similar to the above study
of the critical current. Comparisons between the predicted expression and
results of numerical simulations are shown in figure \ref{fig:fig5}a, which displays the
largest magnitude of the range in dc current for which phase-locking is
observed as a function of the characteristic ratio, $\frac{I_{ac}}{\Omega}$, for
different values of $\Omega$ in the interval $1\le\Omega\le3$. Markers
represent results of numerical simulations and solid curves represent the
corresponding predicted results of Eq.~(\ref{eq:Eq_31}). It is obvious that the simulated
parameter sets provide very good overall validation of the perturbation
analysis, with the larger of the simulated frequencies providing better agreement
than the smaller, as expected. However, an observation common to all simulated
frequencies is that large deviations from the expected behavior are found
for parameter values ($\Omega$ and $I_{ac}$) leading to
$J_0(\frac{I_{ac}}{\Omega})\approx0$. The reason for this discrepancy is likely
due to a dynamical instability, which can be explained by the perturbation
analysis above. The average equilibrium position, $\langle\Psi_y\rangle$, of the
variable $\Psi_y$ can be observed from Eq.~(\ref{eq:Eq_26}) if we write
$\Psi_y=\Psi_y^{(0)}+\psi_y$, where $\Psi_y^{(0)}$ is varying slowly in time
(much slower than $\Omega$), and $\psi_y$ represents all high frequency (including
$\Omega$) contributions ($|\psi_y|\ll1$).
The slow evolution of equation (\ref{eq:Eq_26}) can then be written,
\begin{eqnarray}
2\dot{\Psi}_y^{(0)}+J_0\left(\frac{I_{ac}}{\Omega}\right)\sin\Psi_y^{(0)} & = & 0 \; . \label{eq:Eq_38}
\end{eqnarray}
Thus, we find that the stable position of $\Psi_y$ is:
\begin{eqnarray}
\langle\Psi_y\rangle & = & \left\{\begin{array}{ccc} 0 & , &
J_0\left(\frac{I_{ac}}{\Omega}\right) > 0 \\
\pi & , & J_0\left(\frac{I_{ac}}{\Omega}\right) < 0 \end{array}\right. \; . \label{eq:Eq_39}
\end{eqnarray}
The consequence of this abrupt transition in $\Psi_y$ is that the locking
phase, $\Phi_x^{(0)}$, must experience a similar abrupt transition of $\pi$,
as can be seen from equation (\ref{eq:Eq_11}). We therefore claim that the apparent discrepancy
observed between the numerical simulations and the perturbation theory near
the roots of $J_0\left(\frac{I_{ac}}{\Omega}\right)$, is a result of dynamical
instabilities arising from switching the average phase, $\langle\Psi_y\rangle$
between $0$ and $\pi$.

We finally show the comparisons of the center of the phase-locked step
as a function of the characteristic ratio, $\frac{I_{ac}}{\Omega}$, for
different values of $\Omega$ in the interval $1\le\Omega\le3$. The predicted
behavior, Eq.~(\ref{eq:Eq_32}), is subject to the same issues as the predicted range of
phase-locking since the origin of both expressions is Eq.~(\ref{eq:Eq_30}). Figure \ref{fig:fig5}b shows
the offset, $\Delta I_1=I_1-\Omega$, between the center of the step, $I_1$, and the Ohmic
curve. As is the case for the phase-locking range shown in figure \ref{fig:fig5}a, the
comparison between numerical simulations (markers) and the corresponding
predicted offsets
(solid curves) is very good, except for the instabilities near the roots of
$J_0\left(\frac{I_{ac}}{\Omega}\right)$. 
Notice that the phase-locking analysis leading to the predictions
Eqs.~(\ref{eq:Eq_31}) and (\ref{eq:Eq_32}) does not depend on the sharp $\Psi_y$ transition between $0$ and $\pi$.
The reason is
that this transition provides only a sign change in the effective equations of
phase-locking, and the magnitudes of locking range and offset are therefore unaffected
as long as $J_0\left(\frac{I_{ac}}{\Omega}\right)\neq0$.

Based on the above presented comparisons between the numerical simulations
of critical currents, range of phase-locking and position of the phase-locked
step in the $IV$ characteristics of a transversely ac-driven JJA and the corresponding
results from simple perturbation analysis, we conclude that the high frequency
behavior is well described by the presented analytical treatment.

\subsection{Critical Current and Phase-Locking for intermediate and low $\Omega$}

\begin{figure}[tbp]
\centerline{\epsfxsize=8.5cm \epsfbox{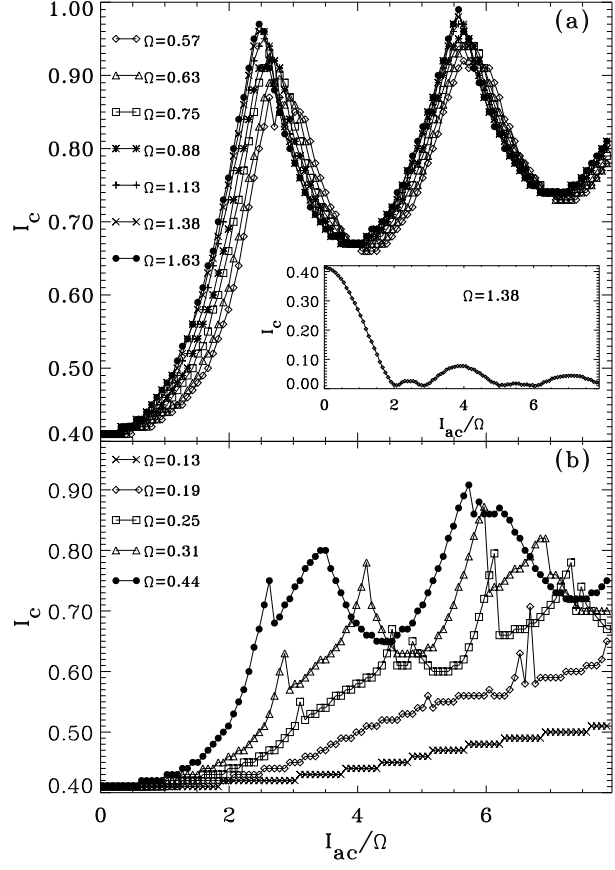}}
\caption{Critical current, $I_c^\uparrow$, as a function of ac-amplitude and frequency,
$I_{ac}/\Omega$ for intermediate frequencies, $\Omega > 0.5$
(a) and for low frequencies, $\Omega < 0.5$ (b).
Inset: Comparison with longitudinal ac-drive for $\Omega=1.38$.
 }
\label{fig:fig6}
\end{figure}

In Fig.~\ref{fig:fig6} we show the critical current
behavior for intermediate and low frequencies.
For intermediate frequencies (figure \ref{fig:fig6}a)
we observe how the critical current, $I_c^\uparrow$,
increasingly deviates from the high frequency behavior outlined above. Even so,
we notice that the overall behavior of the critical current is qualitatively well
described by the analysis leading to equation (\ref{eq:Eq_21}) for $\Omega\ge\frac{1}{2}$.
We have, for comparison, included an example of the critical current for the longitudinally
ac-driven JJA as an inset.
Not surprisingly, decreasing the frequency further (see figure \ref{fig:fig6}b)
results in rather large discrepancy between the high frequency analysis of section
IIIA and the numerical simulations, and no universal behavior of the critical
current as a function of the characteristic ratio, $\frac{I_{ac}}{\Omega}$, can be found.
However, we do observe that the critical current does seem to increase quadratically
for small $I_{ac}$.

\begin{figure}[tbp]
\centerline{\epsfxsize=8.5cm \epsfbox{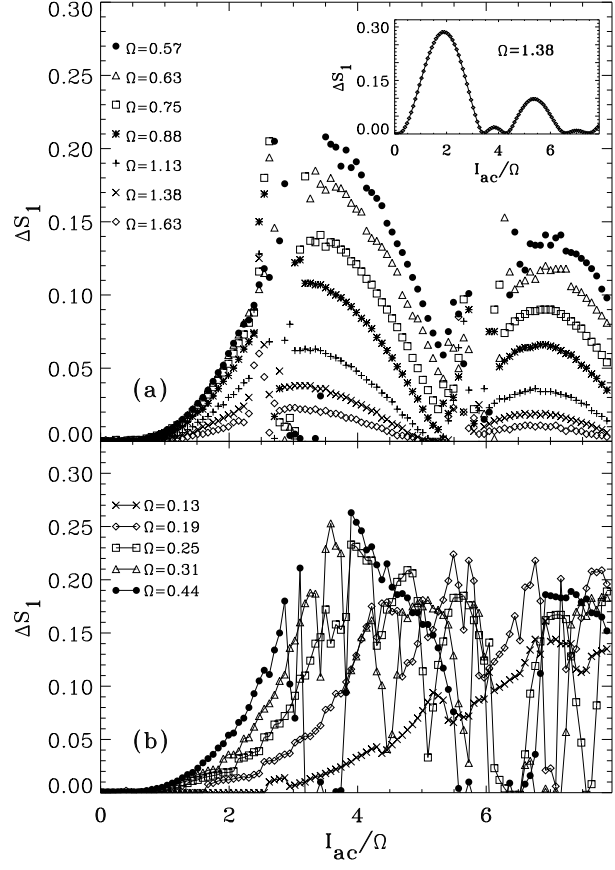}}
\caption{First integer step width, $\Delta S_1$,  vs ac-amplitude and
frequency, $I_{ac}/\Omega$ for intermediate frequencies, $\Omega > 0.5$ (a) and low
frequencies, $\Omega < 0.5$ (b).  Inset: Comparison with longitudinal ac-drive for
$\Omega=1.38$.
}
\label{fig:fig7}
\end{figure}

In Fig.~\ref{fig:fig7} we show the phase-locking range, $\Delta S_1$, at $v_x=\Omega$
as a function of $I_{ac}/\Omega$ for intermediate ($\Omega > 0.5$)
and low frequencies ($\Omega < 0.5$). Again, as for the critical current
we observe that the intermediate frequency range provides for reasonably good
qualitative comparisons between numerical simulations and the high frequency analysis
of section IIIA.
We have, for comparison, included an example of the comparable range of phase-locking
for the longitudinally ac-driven JJA as an inset.
A noticeable feature of figure \ref{fig:fig7}a is that the dynamical
instability discussed above around $J_0\left(\frac{I_{ac}}{\Omega}\right)=0$ seems to
widen as the frequency is lowered.
Figure \ref{fig:fig7}b shows how this instability provides for increasing discrepancy
between high frequency analysis and numerical simulations. However, we notice that
even the very low frequencies retain the basic feature of quadratic growth of
the phase-locking range as a function of $I_{ac}$ for small $I_{ac}$.

\subsection{Dynamics of Phase-Locking}

\begin{figure}[tbp]
\centerline{\epsfxsize=8.5cm \epsfbox{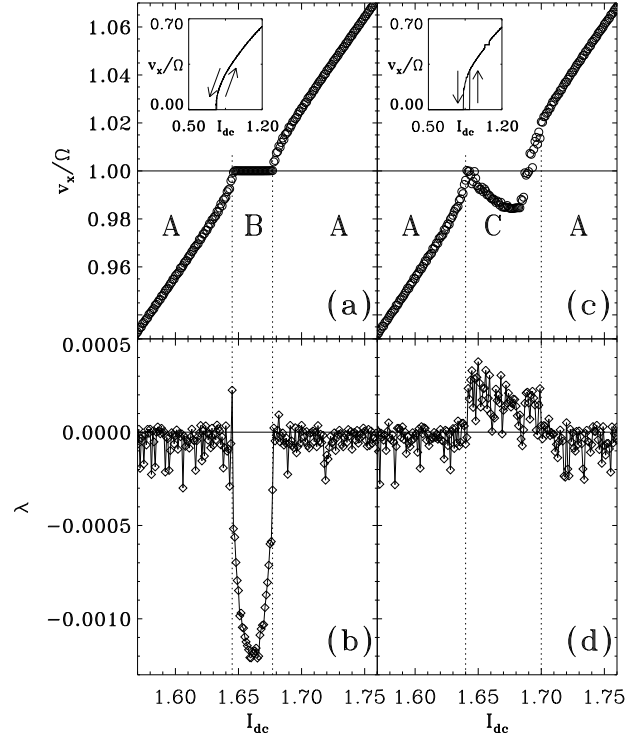}}
\caption{ $IV$ curves  for $\Omega=1.5$ and their corresponding Lyapunov
 exponents, $\lambda$: (a) Part of  $IV$ curve with phase locking step for 
 $I_{ac}=3.0$.
 Inset: detail of  $IV$  curve  near the critical current. No
 hysteresis in $I_c$ is observed. (b) $\lambda$ for $I_{ac}=3.0$.
 (c) $IV$ curve for $I_{ac}=4.05$, no phase locking step is observed.
 Inset: detail of $IV$ near the critical current:
 hysteresis in $I_c$. (d) $\lambda$ for $I_{ac}=4.05$.
 }
\label{fig:fig9}
\end{figure}

Let us now analyze in detail the dynamics of the
voltage responses (increasing and
decreasing dc-current) to elaborate on our previous results:  critical
current hysteresis and  windows without transverse phase locking.
We calculate $IV$ curves and Lyapunov exponents as a function of
$I_{ac}$ and $\Omega$.
In order to distinguish between periodic or quasi-periodic dynamics and
chaotic dynamics  we calculate  the maximum Lyapunov exponent,
$\lambda$, following the standard methods of nonlinear dynamics.\cite{strog,liap}
This means that a small perturbation, $\vec \epsilon(0)$,
to the initial condition will displace the new trajectory
by an amount $|\vec \epsilon (t)|\sim |\vec\epsilon(0)|e^{\lambda t}$.
The Lyapunov exponent is then defined as
$$ \lambda = \lim_{t\rightarrow\infty} \frac{1}{t}\ln{\frac{|\vec
\epsilon (t)|} {|\vec\epsilon(0)|}} = \lim_{t\rightarrow\infty}
\lambda(t)\;.$$
To recognize a chaotic trajectory we evaluate the maximum Lyapunov exponent.
If $\lambda > 0$ the trajectory is locally unstable; i.e.,
initial points that are arbitrarily close to each other are macroscopically
separated by the flow after a sufficiently long time and the  attractor is
chaotic.  Negative Lyapunov exponents are obtained when
trajectories that start sufficiently close to a subset are attracted to it.
Here we will show two particular cases:
$I_{ac}/\Omega=3.0/1.5 = 2$, corresponding to a set of parameters where
no hysteresis is obtained, and $I_{ac}/\Omega=4.05/1.5 = 2.7$,
corresponding to the hysteretic regime.
In Fig.~\ref{fig:fig9} we plot the $IV$ curves and maximum Lyapunov exponents for
$\Omega=1.5$ and $I_{ac}=3.0$ (a-b) and $I_{ac}=4.05$ (c-d). The
exponents are estimated from $\lambda\approx\lambda(t)$ 
after a finite time  $t=1024T$, with $T=2\pi/\Omega$.
For $I_{ac}=3.0$ we show a range in $I_{dc}$ where a wide transverse phase
locking step exists (Fig.~\ref{fig:fig9}a), 
and the corresponding maximum Lyapunov exponent, $\lambda$, 
is shown in Fig.~\ref{fig:fig9}b.
We see that within the step we have $\lambda <0$, with the most
negative value of $\lambda$ at the center of the step. Outside
the steps, the Lyapunov exponent is nearly equal to zero,
$\lambda\lesssim0$, corresponding to quasiperiodic behavior.
A different behavior is obtained for $I_{ac}=4.05$, shown in
Fig.~\ref{fig:fig9}c, where we see that the step disappears and thus there
is no transverse phase locking in the same  $I_{dc}$ range where we find a
step in Fig.~\ref{fig:fig9}a.
The maximum Lyapunov exponent, plotted in Fig.~\ref{fig:fig9}d, is
small but positive for the $I_{dc}$ range around the expected location of 
the step, the smallness of $\lambda$  implies that the dynamics
can be either  chaotic ($\lambda >0$) or quasiperiodic ($\lambda=0$) 
in this case.
The $IV$ curves near the corresponding critical currents are shown as insets.
We see that the absence of  hysteresis in
critical current is associated with the occurrence of
transverse phase locking. Inversely, hysteresis in critical
current  is obtained for approximately the same parameters for which
transverse phase locking is absent.
This is in agreement with the above analysis that indicates the critical
current hysteresis is present in the vicinity of
$J_0\left(\frac{I_{ac}}{\Omega}\right)=0$, which is also the location of the
dynamical instabilities of the locked phase of the $\Delta S_1$ step.

\begin{figure}[tbp]
\centerline{\epsfxsize=8.5cm \epsfbox{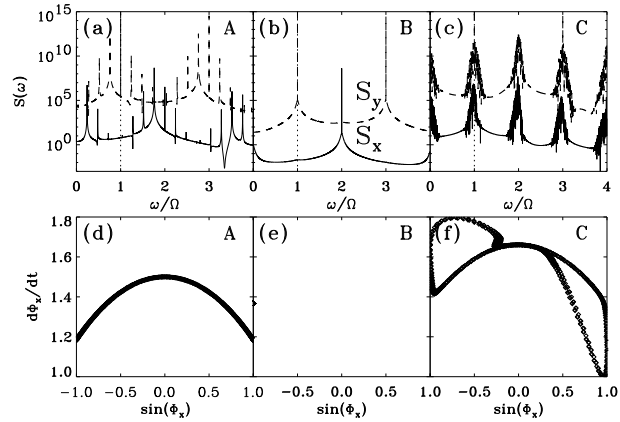}}
\caption{ Voltage power spectra and Poincar\'e sections
  for $\Omega=1.5$
 in different $I_{dc}$ regimes: A, B and C
 (see Fig.~\ref{fig:fig9}. 
 (a) Transverse, $S_y$, and longitudinal, $S_x$, power spectra for
$I_{dc}=1.5$ and $I_{ac}=3.0$. A regime.
(b) $I_{dc}=1.66$ and $I_{ac}=3.0$. B regime.
(c) $I_{dc}=1.66$ and $I_{ac}=4.05$. C regime.
(d),(e) and (f) are the corresponding Poincar\'e sections.
Power spectrum $S_x$ is plotted displaced in the $y$-axis for clarity.
}
\label{fig:fig10}
\end{figure}

In summary,  around the transverse phase locking step region
we can distinguish three different voltage responses: A, B and C,
which are indicated in Fig.~\ref{fig:fig9}. 
We now calculate the voltage power spectrum and Poincar\'e sections
to distinguish these  three types of dynamical behavior. This way to
characterize dynamical behaviors  was used before in capacitive
rf-biased JJA. \cite{kautz, thomas} We analyze both transverse   and
longitudinal voltage power spectra.  From the instantaneous transverse
voltage we obtain the transverse voltage power spectrum:
\begin{equation}
 S_y (\omega)=\Bigl |\frac{1}{T_t} \int_0^{T_t} dt
V_y(t) \exp (i\omega t) \Bigr|^2 \; ,
\end{equation}
where $T_t=N_t \Delta t$. From the instantaneous longitudinal voltage,
$V_x$, we obtain the longitudinal voltage power spectrum:
\begin{equation}
S_x (\omega)=\Bigl |\frac{1}{T_t} \int_0^{T_t} dt V_x(t)  \exp (i\omega t)
\Bigr |^2 \; .
\end{equation}
For studying the nature of the attractor in the
different regimes  it is useful to consider a Poincar\'e section of the
phase-space trajectories. \cite{strog}
We consider the stroboscopic
Poincar\'e section of the trajectories in the $d\Phi_x/dt$
vs.\ $\sin\Phi_x$ plane, recording the values taken by these variables
each period of the ac drive.
In Fig.~\ref{fig:fig10} we show the power spectra and Poincare sections
for $\Omega=1.5$ and for the
$I_{ac}$ and $I_{dc}$ values  corresponding to the A, B and C
regimes. For each case we show the longitudinal, $S_x(\omega)$, and transverse,
$S_y(\omega)$, voltage power spectra as a function of $\omega / \Omega$ and
their corresponding Poincar\'e sections.
Let us first discuss the case corresponding to the regime B, in which
there is transverse phase-locking. This is shown
in Fig.~\ref{fig:fig10}b for $I_{dc}=1.66$ and $I_{ac}=3.0$,
which corresponds to the step with mean voltage
$v_x=\langle V_x\rangle=\Omega$ (see Fig.~\ref{fig:fig9}a,  regime B).
We see that the longitudinal power spectrum, $S_x(\omega)$,
presents a delta-like peak for $\omega = 2\Omega$.
Thereby,
the first harmonic of longitudinal voltage fluctuations is locked to
$2\Omega$, as expected for this step, since it corresponds to $n=2$ in
the phase-locking condition $\omega_0=n\Omega$.
The phase-locking with a double
frequency corresponds to the case when the vortex lattice oscillates in
full synchrony with the transverse ac-current,
and the ground state repeats itself after one period of the ac drive.
In the transverse voltage power spectrum,
$S_y$, there is a sharp  peak  at  $\Omega$.
This is characteristic of transverse phase-locking: the dynamics
in the transverse direction locks at half the frequency than
the dynamics of the longitudinal direction. This is so because
in a single period of the ac-drive, $T=2\pi/\Omega$,
the longitudinal component moves forward $n$ steps in the
lattice period $a$, while the transverse component completes only
the first half of its oscillation.
In Fig.~\ref{fig:fig10}e, we show the Poincar\'e section corresponding to
this case in the regime B.
The figure shows a very localized Poincar\'e section since the trajectory
always comes back approximately to the same location in phase space in 
each ac cycle,
since the trajectory is periodic (closed orbit).

Now we analyze the case corresponding to the A regime, which is for a
current outside the step, $I_{dc}=1.5$, see Fig.~\ref{fig:fig9}a. 
In this case we see again in $S_x(\omega)$ a peak at $2\Omega$
and in $S_y(\omega)$  a peak at $\Omega$. However, the peaks now
have a small broadening, and small amplitude satellite peaks have appeared
at neighboring frequencies.
This is evidence of another kind of long-term
behavior, namely quasiperiodic dynamics. We can corroborate this with the
corresponding Poincar\'e section shown in Fig.~\ref{fig:fig10}d.
It consists  now on a closed one dimensional
curve, which means that trajectories wind around
on a torus, never intersecting itself and yet never quite closing,
typical of a quasiperiodic orbit.
We have also looked at the time dependent estimates of the
Lyapunov exponent, $\lambda(t)$. We find that $\lambda(t) < 0 $ for 
finite $t$, but its absolute value tends to zero for long times as 
$1/t$, consistent with quasiperiodic behavior.

Let us now study the last case, corresponding to regime C.
 This is done for
$I_{dc}=1.66$ and $I_{ac}=4.05$ in Fig.~\ref{fig:fig10}c.
We see that there are broad peaks in the
spectrum in both directions, $S_x(\omega)$ and $S_y(\omega)$,
and that there is a marked  increase in the power spectra for low
frequencies. 
Moreover, in Fig.~\ref{fig:fig10}f we show the corresponding Poincar\'e
section which
consists on successive points jumping from one region of
phase space to
another and  forming a complex curve, which does not seem to close on
itself. It is rather difficult to decide from this plot if it
corresponds to a quasiperiodic orbit or to a low dimensional attractor
of a weakly chaotic orbit.
We have obtained also the time dependent estimate of the Lyapunov exponent
also for this case. We find that $\lambda(t) > 0 $ for all $t$, but
its magnitude is decreasing with time as $1/t$ as far as we have been
able to observe. The fact that $\lambda(t)$ is possitive for finite
$t$ means that there is a dynamical instability that  causes a 
seemingly chaotic behavior at intermediate times  and a large noise
as seen in the low frequency power spectrum. However, for long times
it is very likely that the system will settle in a quasiperiodic
dynamics with $\lambda(t)\rightarrow0$. In any case, the regime C is
very different from the regime A, as can be seen by comparing
the power spectra of Fig.~\ref{fig:fig10}a and Fig.~\ref{fig:fig10}c.

Another view of the dynamics can be obtained by looking at the behavior of the 
Lyapunov exponent and the noise in the region, where a step is expected, 
as a function of $I_{ac}/\Omega$.
We proceed as follows: for a given $I_{ac}, \Omega$,
we compute the set of values of Lyapunov exponents $\lambda$ and 
low frequency longitudinal noise
$S_0 = \lim_{\omega\rightarrow0} S_x(\omega)$  that correspond
to currents $I_{dc}$ in the  region of voltage where a step is 
expected. (We look at $I_{dc}$ values for which 
$\Omega-\epsilon < V_x < \Omega + \epsilon$, we consider
$\epsilon=0.005$). 
We plot the resulting set of values of $\lambda$ as a function
of $I_{ac}/\Omega$ in Fig.~10a and  the values of $S_0$
as a functions of $I_{ac}/\Omega$ in Fig.~10b. 
The vertical lines in the plot correspond to
the zeroes of $J_0(I_{ac}/\Omega)$. We see clearly that near these values
there are windows of dynamical instability where $\lambda\gtrsim 0$ 
and where the noise $S_0$ is large. In the regions of phase-locking 
we find a couple of interesting results that are worth mentioning. 
(i) The most negative
value of the Lyapunov exponent occurs in the middle of the phase-locked step and
its magnitude is proportional to the step width $\Delta S_1$, as given
by Eq.~(\ref{eq:Eq_31}). (ii) The largest value of the noise $S_0$
occurs at the edge of the phase-locked step; its magnitude is also 
proportional to the step width $\Delta S_1$, as given
by Eq.~(\ref{eq:Eq_31}). 

\subsection{Results for Large JJA}

We will now consider the quality of the simple $2\times2$ model as
representing the dynamics of large $N\times N$ JJAs.
It is known that collective effects at high currents
may  come into play.   At high currents, the $Z_2$ symmetry of the
ground state  can be broken because
a driving current can induce domain walls. \cite{mon,niels,marconi2}
Simulations of  $IV$ curves with the RSJ model and free
boundary conditions,  for $f=1/2$ and $T=0$ have reported a chaotic regime at
$I > I_c$  related to the motion of domain walls. \cite{falo}
It has been shown that open boundary conditions nucleate domain walls
leading to a critical current lower than de analytic value
$I_c=0.35 < \sqrt{2}-1$ at $T=0$.\cite{minn}
Moreover, Ciria and Giovanella \cite{ciria} have shown microscopically that
different dynamical states are possible for the longitudinal Shapiro steps.
Besides the
checkerboard ground state configuration, other stable solutions with
domain-walls are possible.
Then, depending on dc current value and history,
domain walls can appear, which  are not permitted in the four plaquette model.
Therefore,
in order to evaluate to what extent the four plaquette model is valid
in the transverse ac driven case,
we have calculated numerically $IV$ curves for $N\times N$ arrays, for
$N=8,16,32,64$, with the full RSJ model used before in
Ref.~\onlinecite{marconi,marconi2}. We use periodic boundary conditions in both
directions in the presence of an external dc current, $I_{dc}$, plus a
perpendicular  ac current, $I_{ac} \sin(\Omega t)$.  We solve the dynamical
equations  with time step $\Delta t = 0.1 \tau_J$  ($\tau_J=2\pi c R_N
I_0/\Phi_0$) and total integration time $t_{\rm int}=2^{15}\Delta t$ 
after a transient $t_{\rm int}/2$.
We calculate $IV$ curves as a function of $I_{ac}$ and $\Omega$, increasing dc
current, $I_{dc}^\uparrow$, from  checkerboard ground state at
$I_{dc}=0$ and then decreasing dc current, $I_{dc}^\downarrow$, from the phase
configuration obtained at high current.
We use a dc current step $\Delta I_{dc}=0.01$ to obtain $I_c$ and 
$\Delta I_{dc}=0.0001$ to calculate the step width.

\begin{figure}[tbp]
\centerline{\epsfxsize=8.5cm \epsfbox{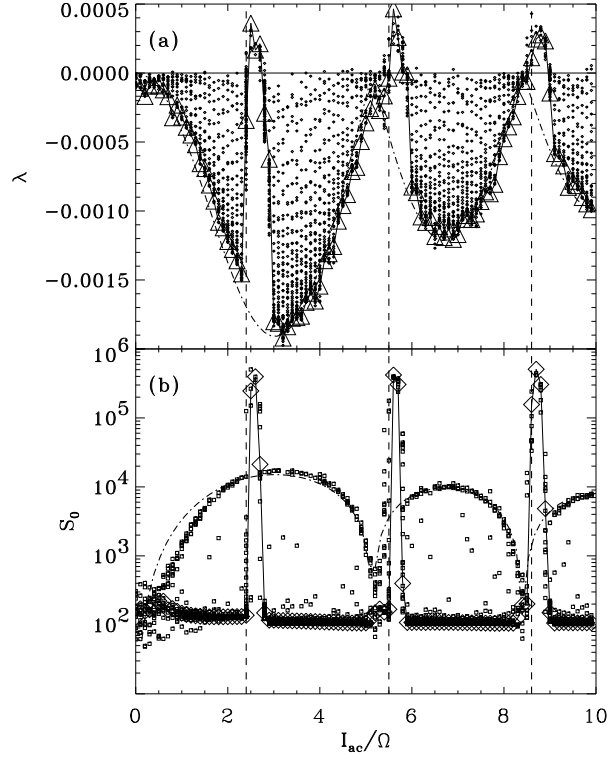}}
\caption{Lyapunov exponents $\lambda$ and low frequency noise $S_0$ for
currents $I_{dc}$ giving voltages near $V=\Omega$ plotted as a function
of $I_{ac}/\Omega$. Vertical dashed lines correspond to the zeros of 
$J_0(I_{ac}/\Omega)$.
 (a) Lyapunov exponents. Symbol $\triangle$ indicates value
of $\lambda$ at the center of the phase-locked step. Dot-dashed line:
curve proportional to $\Delta S_1 (I_{ac}/\Omega)$ as given
by Eq.~(\ref{eq:Eq_31}).
(b)Low frewquency noise. Symbol $\diamond$ indicates value
of $S_0$ at the center of the phase-locked step. Dot-dashed line:
curve proportional to $\Delta S_1 (I_{ac}/\Omega)$ as given
by Eq.~(\ref{eq:Eq_31}).  }
\label{fig:fig11}
\end{figure}

One of the relevant results with the four plaquette model is the
dependence of the critical current with $I_{ac}/\Omega$  for high
frequencies, as shown in Fig.~\ref{fig:fig4}a and Fig.~\ref{fig:fig6}a.
We have also calculated
$I_c$ as a function of  $I_{ac}/\Omega$ for high $\Omega$ in large JJA
arrays. In Fig.~\ref{fig:fig11}a we show the case for a 
particular high frequency
value  in a $32 \times 32$ array.  We see that it has the same behavior
as observed in the four plaquette model: $I_c(I_{ac},\Omega)\ge
I_c(0,0)$,  ranges of  $I_{ac}/\Omega$  around the maxima of $I_c$
where there is hysteresis, and a quadratic   increase   with $I_{ac}$
for $I_{ac}/\Omega \ll 1$.  We compare with the analytical results 
expected for $I_c^\uparrow$, Eq.~(\ref{eq:Eq_21}) 
and $I_c^\downarrow$, Eq.~(\ref{eq:Eq_37}), which are represented
by dot-dashed lines.  We see that $I_c^\uparrow$
obtanained numerically for a large array is in excellent agreement
with the analytical result for the $2\times2$ system. This is quite
reasonable, since $I_c^\uparrow$ corresponds to the limit of stability
of the checkerboard ground state, which is well represented by
the $2\times2$ model. On the other hand, the  $I_c^\downarrow$
shows some small deviation from the $2\times2$ result,
$I_c^\downarrow(N\times N)\lesssim I_c(2\times2)$. Also 
the range in $I_{ac}/\Omega$ where there is hysteresis is bigger in
a large system.  
The current $I_c^\downarrow$ corresponds to the low current limit of 
stability of the moving (non-zero voltage) state. 
In large systems, the moving state can have domain walls, as was
found in Ref.~\onlinecite{niels,marconi2}, and the presence of domain walls
can lead to  a lower $I_c^\downarrow$.

In order to analyze more quantitatively in
which $I_{ac}/\Omega$ ranges  the collective effects could be more
relevant,  we focus on two cases: case $a$ corresponding to
values that do not show  hysteresis  in the critical current in a small
system, but are close to the edge of the $I_{ac}/\Omega$ range of hysteresis, 
and
case $b$ corresponding to values  that show hysteresis in the $2\times2$
system. For each case
we calculate the critical current both increasing and decreasing the dc
drive, and therefore they correspond to  $a \uparrow$, $a \downarrow$,
$b \uparrow$ and $b \downarrow$ in Fig.~\ref{fig:fig11}a. 
We show in the inset of
Fig.~\ref{fig:fig11}a the critical currents obtained for all these cases
as a function of system size, $N$.  In case $a$, corresponding to the
non-hysteretic region,  we see that there is no size effect in
$a\uparrow$ up to $N=64$. Also we see that $a\downarrow=a\uparrow$ for
$N\le 32$, while for $N=64$ we find that  hysteresis 
has appeared and $a\downarrow<a\uparrow$.
In the hysteretic region, case $b$,  size dependent critical
currents  are obtained for $b\downarrow$, while $b\uparrow$ is size
independent. Moreover, the amplitude of the hysteresis, $b\uparrow-b\downarrow$
weakly increases with  system size.

In Fig.~\ref{fig:fig11}b we show the range of phase locking $\Delta S_1$
as a  function of $I_{ac}/\Omega$ for the $32\times32$ array.
We find that, when there is phase-locking, 
the numerically obtained   $\Delta S_1$ is very
accurately described by the analytical result  of Eq.~(\ref{eq:Eq_31}) 
for the $2\times2$ model. In the inset of Fig.~\ref{fig:fig11}b
we show the size dependence of $\Delta S_1$ for the case marked as
$c$ in the plot (it corresponds to the same $I_{ac}/\Omega$ of
case $a$ of Fig.~\ref{fig:fig11}a). There is no appreciable size
dependence. As observed in the simulations of the $2\times2$ system,
we also find here that 
the phase-locking is lost near the zeros of $J_0(I_{ac}/\Omega)$ due
to the presence of dynamical instabilities. Also we observe that the
presence of hysteresis in the
critical current is nearly coincident with the absence of phase locking.  
We find that with increasing system size these  regimes of dynamical instability 
are amplified in their extension both in their $I_{dc}$ dependence and in 
their range of $I_{ac}/\Omega$ around the zeros of $J_0(I_{ac}/\Omega)$.  
This means that
the dynamical instabilities  detected in the four plaquette system
can lead to an increased spatiotemporal chaos in larger systems
where collective effects are  important.

\section{CONCLUSIONS}

It is important to
point out that there are no trivial connections between  vortex dynamics in
the fully frustrated JJA and that of a commensurate  vortex lattice moving in
a rectangular pinning potential in a bulk superconductor. \cite{reichhardt2}
This is so because the London model with vortices interacting through pair
potentials apply to JJAs only in the limit of very low vortex density,
such that $f=Ha^2/\Phi_0 \ll 1$. \cite{marconi} The fully frustrated  case
represents, in this last respect, an interesting limit for studying,
where the complete phase field, rather than just the positions
of vortices, should be taken into account to describe the dynamics.

We have found transverse phase locking steps in fully frustrated JJA. This
new type of (fractional) giant phase-locking steps presents marked differences with
the well known longitudinal fractional giant Shapiro steps. Particularly,
the presence of the transverse ac force increases the
critical depinning current with respect to the case without ac drive
(or with a longitudinal ac drive). We have analyzed both analytically
and numerically the behavior
of the steps as a function of ac amplitude $I_{ac}$ and frequency
$\Omega$. For $I_{ac}/\Omega \ll 1$,
the depinning critical current and the phase-locked step width
$\Delta S_1$ for $V=\Omega \hbar/2e$,
increase quadratically with $I_{ac}$. For $I_{ac}/\Omega > 1$ we have found
windows of $I_{ac}/\Omega$ where depinning is hysteretic and phase
locking is destroyed due to dynamical instabilities. 
The emergence of a weakly chaotic behavior
at zero temperature, in a system with non capacitive junctions, is another
particular characteristic of transverse phase locking which is absent in
longitudinal phase locking in overdamped JJA. 
Comparing with the behavior of large fully
frustrated arrays we have found that transverse phase locking can be
well described by an effective four plaquette model, and that
collective effects become more important close to
the regions of dynamical instability of the four-plaquette model. Our results
could be observed experimentally in JJA. In particular,
the enhancement of the critical depinning current with a transverse
ac drive could be an interesting experimental consequence of the phenomena
reported here.

\acknowledgments

We acknowledge discussions with G. Carneiro, H. Pastoriza, C. Reichhardt
and D. Shalom. 
Parts of this work were supported by the Director, Office of
Advanced Scientific Computing Research, Division of Mathematical, Information
and Computational Sciences of the U.S.~Department of Energy under
contract number DE-AC03-76SF00098. The work in Argentina was supported
by CNEA, Conicet and ANPCyT (PICT99-03-06343).

\newpage

\begin{figure}[tbp]
\centerline{\epsfxsize=8.5cm \epsfbox{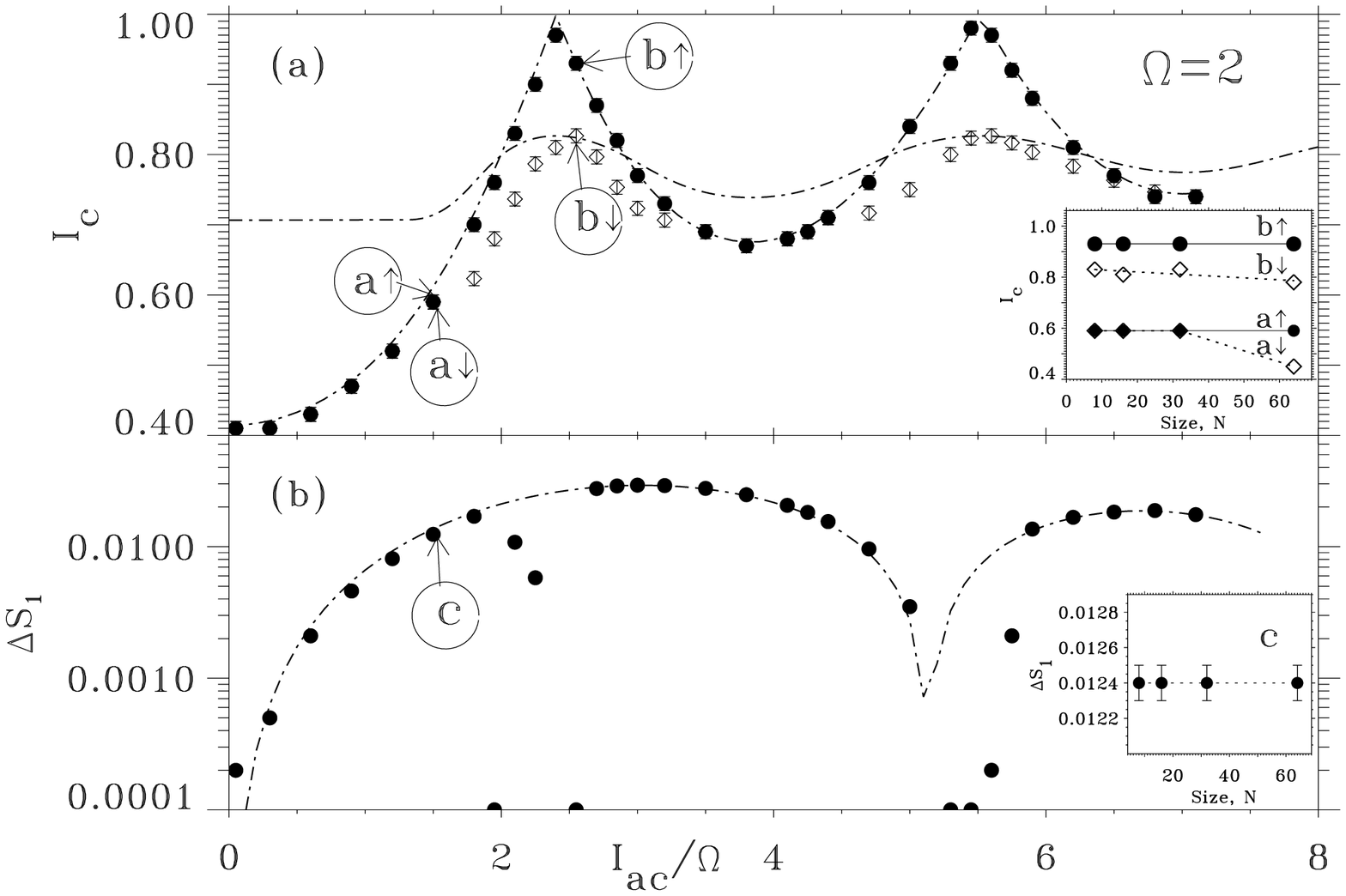}}
\caption{Critical currents $I_c$ and step witdths $\Delta S_1$ 
obtained from numerical simulation in large arrays
($32 \times 32$ junctions) as a function of $I_{ac}/\Omega$.
(a)  $I_c$ obtained
increasing $I_{dc}$  ($I_{c}^{\uparrow}$, $\bullet$) and decreasing
$I_{dc}$ ($I_{c}^{\downarrow}$, $\diamond$). Dot-dashed lines show the
analytical results of Eq.~(\ref{eq:Eq_21}) and Eq.~(\ref{eq:Eq_37}).
Inset shows the size dependence of $I_{c}^{\uparrow}$ and $I_{c}^{\downarrow}$
for system of size $N\times N$, corresponding to the cases $a$ and $b$
indicated in the plot.
(b) Width of the first integer phase-locked step $\Delta S_1$, obtained
numerically for a $32\times32$ array: $\bullet$, and analytical result
of Eq.~(\ref{eq:Eq_31}): dot-dashed line. Inset shows size dependence
of  $\Delta S_1$ for the case $c$ shown in the plot.} 
\label{fig:fig12}
\end{figure}


\begin{thebibliography}{}

\bibitem{bak}{Per Bak, Physics Today, December 1986, 36.}

\bibitem{shapiro}{ S. Shapiro, Phys. Rev. Lett. {\bf 11}, 80 (1963).}

\bibitem{barone}{
A. Barone and G. Paterno, {\it Physics and applications of the Josephson
Effect} (Wiley, New York, 1982).}

\bibitem{leeman}{ Ch. Leeman, Ph. Lerch, and P. Martinoli, Physica
{\bf 126B}, 475 (1984).}

\bibitem{expfgs}{S. P. Benz, M. S. Rzchowski,
M. Tinkham, and C. J. Lobb, Phys. Rev. Lett. {\bf 64}, 693 (1990);
Physica B{\bf 165-166}, 1645 (1990).}

\bibitem{expshap}
{H.\ C.\ Lee,  R.\ S.\ Newrock, D.\ B.\ Mast,
S.\ E.\ Hebboul, J.\ C.\ Garland, and C.\ J.\ Lobb,  Phys. Rev. B {\bf 44},
921 (1991); S.\ E.\ Hebboul and J.\ C.\ Garland,  Phys. Rev. B {\bf43},
13703 (1991); Phys.\ Rev.\ B {\bf 47}, 5190 (1993).}

\bibitem{lee}{K. H. Lee, D. Stroud, and J. S. Chung, Phys. Rev. Lett.
{\bf 64}, 962 (1990); K. H. Lee and D. Stroud,
Phys. Rev. B {\bf 43}, 5280 (1991).}

\bibitem{simff}{J.U. Free, S. P. Benz, M. S. Rzchowski,  M. Tinkham,
and C. J. Lobb and M. Octavio,  Phys. Rev. B {\bf 41}, 7267 (1990).}

\bibitem{fourp}{ M.S. Rzchowski, L.L. Sohn, and M. Tinkham,
Phys. Rev. B {\bf 43}, R8682 (1991).}

\bibitem{octav}{ M. Octavio, J.U. Free, S.P. Benz, R.S. Newrock,
D.B. Mast, C.J. Lobb, Phys. Rev. B {\bf 44}, 4601 (1991).}

\bibitem{ciria} {J. C. Ciria and C. Giovanella, J. Phys. Cond. Matt.
{\bf 8}, 3057 (1996); J. C. Ciria and C. Giovanella,
in {\it Macroscopic Quantum Phenomena and Coherence in Superconducting
Networks}, Ed. C. Giovannella and M. Tinkham (World Scientific,
Singapore, 1995), p.301.}

\bibitem{acvs} {D.\ Dom\'{\i}nguez, J. V. Jos\'{e}, A. Karma and
C. Wiecko,  Phys.\ Rev.\ Lett.\  {\bf 67}, 2367 (1991);
D.\ Dom\'{\i}nguez  and J.\ V.\ Jos\'{e},
Phys.\ Rev.\ Lett.\  {\bf  69}, 514 (1992);
 Phys. Rev. B {\bf 48}, 13717 (1993),
and Int. J. Mod. Phys. B {\bf 8}, 3749 (1994).}

\bibitem{giov} {C. Giovanella, F. Ritort and A. Gianelli,
Europhys. Lett. {\bf 29}, 419 (1995).}


\bibitem{vanlook}{ L. Van Look, E. Rosseel, M. J. Van Bael, K. Temst,
V. V. Moshchalkov, and Y. Bruynseraede, Phys. Rev. B {\bf 60}, R6998 (1999).}

\bibitem{reichhardt}{C. Reichhardt, R. T. Scalettar, G.T. Zim\'anyi, and
N. Gr{\o}nbech-Jensen, Phys. Rev. B {\bf 61}, R11914 (2000)}

\bibitem{martinoli}{ P. Martinoli, O. Daldina, C. Leemann, and E. Stocker,
Solid State Commun. {\bf 17}, 205 (1975).}

\bibitem{kokubo}{N. Kokubo, R. Besseling, V. M. Vinokur and P. H. Kes,
Phys. Rev. Lett. {\bf 88}, 247004 (2002).}

\bibitem{gruner}{ G. Gr\"uner, Rev. Mod. Phys. {\bf 60}, 1129 (1988);
S. Bhattacharya, J. P. Stokes, M. J. Higgins,  and R. A. Klemm,
Phys. Rev. Lett. {\bf 59}, 1849 (1987); M. J. Higgins, A. A. Middleton,
and S. Bhattacharya, {\it ibid.} {\bf 70}, 3784 (1993); S. N. Coppersmith
and P. B. Littlewood, {\it ibid.} {\bf 57}, 1927 (1986); A. A. Middleton,
O. Biham, P. B. Littlewood, and P. Sibani,  {\it ibid.} {\bf 68}, 1586
(1992).}


\bibitem{fiory}{ A. T. Fiory, Phys. Rev. Lett. {\bf 27}, 501 (1971).}

\bibitem{harris}{ J. M.
Harris, N. P. Ong, R. Gagnon, and L. Taillefer, Phys. Rev. Lett. {\bf 74}, 3684
(1995).}

\bibitem{kolton}{ A. B. Kolton, D. Dom\'{\i}nguez, N. Gr{\o}nbech-Jensen, Phys.
Rev. Lett. {\bf 86}, 4112 (2001).}

\bibitem{reichhardt2}{ C. Reichhardt, A. B. Kolton, Daniel Dom\'{\i}nguez and
N. Gr{\o}nbech-Jensen, Phys. Rev. B {\bf 64}, 134508 (2001);
C. Reichhardt and C. J. Olson, Phys. Rev. B {\bf 65}, 174523 (2002).}

\bibitem{kolton3}{A. B. Kolton, D. Dom\'{\i}nguez, N. Gr{\o}nbech-Jensen,
Phys. Rev. B {\bf 65}, 184508 (2002).}

\bibitem{koshelev}{ A. E. Koshelev and V. M. Vinokur, Phys. Rev. Lett.
{\bf 73}, 3580  (1994); S. Scheidl and V. M. Vinokur, Phys. Rev. B
{\bf 57}, 13800 (1998); T. Giamarchi and P. Le Doussal,
Phys. Rev. Lett. {\bf 76}, 3408 (1996);
P. Le Doussal and T. Giamarchi, Phys. Rev. B {\bf 57}, 11356
(1998); L. Balents, M. C. Marchetti and L. Radzihovsky, {\it ibid.}
{\bf 57}, 7705 (1998).}


\bibitem{moon}{  K. Moon {\it et al.},
Phys. Rev. Lett. {\bf 77}, 2778 (1996); S. Ryu {\it et al.},
{\it ibid.} {\bf 77}, 5114 (1996);
N. Gr\o nbech-Jensen {\it et al.},
{\it ibid.} {\bf 76}, 2985 (1996);
C. Reichhardt {\it et al.}, {\it ibid.} {\bf 78}, 2648
(1997);  D. Dom\'{\i}nguez {\it et al.},
{\it ibid.} {\bf 78}, 2644 (1997);
C. J. Olson {\it et al.},
{\it ibid.} {\bf 81}, 3757 (1998);
D. Dom\'{\i}nguez, {\it ibid.} {\bf 82}, 181 (1999);
A. B. Kolton, D. Dom\'{\i}nguez, and N. Gronbech-Jensen,
     Phys. Rev. Lett. {\bf 83}, 3061 (1999);
H. Fangohr {\it et al.}, Phys. Rev. B {\bf 63}, 064501 (2001);
K. Bassler {\it et al.}, {\it ibid.} {\bf 64}, 224517 (2001).}

\bibitem{olson}{C. J. Olson and C. Reichhardt, Phys. Rev. B {\bf 61},
R3811 (2000); H. Fangohr {\it et al.}, {\it ibid.} {\bf 63}, 064501
(2001).}

\bibitem{marconi}{V. I. Marconi and D. Dom\'{\i}nguez, Phys. Rev. Lett.
{\bf 82}, 4922 (1999); Phys. Rev. B {\bf 63}, 174509 (2001).}

\bibitem{kolton2}{A. B. Kolton, D. Dom\'{\i}nguez, N. Gr{\o}nbech-Jensen,
Phys. Rev. Lett. {\bf 83}, 3061 (1999); A. B. Kolton, D. Dom\'{i}nguez, C.J. Olson, N. Gronbech-Jensen,
Phys. Rev. B {\bf 62}, R14657 (2000).}


\bibitem{niels}  N.~Gr{\o}nbech-Jensen, A.~R.~Bishop, F.~Falo, and P.~S.~Lomdahl,
Phys.~Rev.~B~{\bf 46}, 11149 (1992).


\bibitem{marconi2}{ V. I. Marconi and D. Dom\'{\i}nguez, Phys. Rev. Lett.
{\bf 87}, 017004 (2001).}






\bibitem{xyff}{S. Teitel and C. Jayaprakash, Phys. Rev. B {\bf 27}, 598 (1983),
Phys. Rev. Lett. {\bf 51}, 1999 (1983).}

\bibitem{mon}{K. K. Mon and S. Teitel, Phys. Rev. Lett. {\bf
62}, 673 (1989).}




\bibitem{ccff}{S. P. Benz, M. S. Rzchowski,  M. Tinkham,
and C. J. Lobb, Phys. Rev. B {\bf 42}, 6165 (1990).}



\bibitem{strog}{Steven H. Strogatz, {\it Nonlinear dynamics and Chaos}, Perseus
Books, Cambridge, Massachusetts, 1998.}

\bibitem{liap} {I. Shimada and T. Nagashima, Prog. Theor. Phys. {\bf 61},
1605 (1979).}

\bibitem{kautz}{ R.L. Kautz and R. Monaco, J. Appl. Phys. {\bf 57}, 875 (1985).}

\bibitem{thomas}{Thomas Hagenaars, PhD Thesis (1995), Universiteit Utrecht.}

\bibitem{falo}{ F.\ Falo, A.\ R.\ Bishop, and P.\ S.\ Lomdahl, Phys.\ Rev.\ B
 {\bf 41}, 10983 (1990).}




\bibitem{minn} B. J. Kim and P. Minnhagen, Phys.\ Rev.\ B
 {\bf 61}, 7017 (2000).
 
\end{thebibliography}
\end{document}